\documentclass[preprint,preprintnumbers,nofootinbib]{revtex4-1}
\usepackage{graphicx}
\usepackage{dcolumn}
\usepackage{bm}
\usepackage{amsmath}
\usepackage{amsfonts} 
\usepackage{latexsym}
\usepackage{bbm}
\usepackage{color}
\usepackage{amssymb}
\usepackage{amsthm}
\usepackage{epsfig}
\usepackage{physics}
\usepackage{hyperref}
\usepackage{comment}
\hypersetup{%
setpagesize=false,
 bookmarksnumbered=true,%
 bookmarksopen=true,%
 colorlinks=true,%
 linkcolor=blue,
 citecolor=red,
}

\begin{document}

\preprint{\textbf{OCHA-PP-374}}

\title{A complex singlet extension of the Standard Model and Multi-critical Point Principle}

\author{Gi-Chol Cho$^1$}
\email{cho.gichol@ocha.ac.jp}
\author{Chikako Idegawa$^2$}
\email{c.idegawa@hep.phys.ocha.ac.jp}
\author{Rio Sugihara$^2$}
\email{rio.sugihara@hep.phys.ocha.ac.jp}
\affiliation{$^1$Department of Physics, Ochanomizu University, Tokyo 112-8610, Japan}
\affiliation{$^2$Graduate school of Humanities and Sciences, Ochanomizu University, Tokyo 112-8610, Japan}
\bigskip

\date{\today}

\begin{abstract}
We study the Multi-critical Point Principle (MPP) in a complex singlet scalar extension of the Standard Model (CxSM). The MPP discussed in this study selects model parameters so that two low-energy vacua realized by scalar fields are degenerate. We further note that the MPP may inhibit the electroweak phase transition (EWPT) in a certain class of models where the tree-level potential plays an essential role in its realization. 
Despite that, we show that strong first-order EWPT still occurs even after imposing the MPP to the scalar potential of the CxSM due to the 1-loop corrections by the new scalar boson. We study the allowed parameter space where a mass of the additional scalar is degenerate with that of the Standard Model Higgs boson, which provides a built-in mechanism to circumvent constraints from dark matter direct detection experiments. The parameter space for the non-degenerate scalar scenario is also studied for comparison.

\end{abstract}

\maketitle

\section{Introduction}\label{sec:Intro}
The discovery of the Standard Model (SM) Higgs boson in the LHC experiment was one of the most important contributions to our understanding of particle physics in recent years~\cite{ATLAS:2012yve,CMS:2012qbp}. 
Although the SM is almost complete with this discovery, various astronomical observations, including precise measurements of the cosmic microwave background radiation, indicate the existence of dark matter (DM), which motivates us to search for new physics beyond the SM (BSM). BSM has been vigorously explored by LHC~\cite{ATLAS:2019nkf,CMS:2020gsy} and DM direct detection experiments~\cite{XENON:2020kmp,LZ:2022ufs}, yet no signals have been obtained, placing strong constraints on parameter space. 

A simple extension of the SM, which includes a DM candidate particle, is to add a complex singlet scalar field, known as the complex singlet extension of the SM (CxSM)~\cite{Barger:2008jx,Barger:2010yn,Gonderinger:2012rd,Coimbra:2013qq,Jiang:2015cwa,Chiang:2017nmu,Cheng:2018ajh,Grzadkowski:2018nbc,Chen:2019ebq,Egle:2022wmq}.  In the CxSM, a real part of an additional complex scalar mixes with the SM Higgs boson, while an imaginary part of the scalar would be a DM candidate. 
It is known that when masses of two CP-even scalars appearing in the CxSM are degenerate, a spin-independent cross section of DM with the nucleons ($\sigma_{\mathrm{SI}}$) could be suppressed, satisfying the results of direct-detection experiments~\cite{Abe:2021nih}. 
Due to the orthogonality of the mixing matrix of CP-even scalar bosons,  the Higgs signal strength in this model is nothing different from the SM prediction in the degenerate limit of two scalar masses. Therefore, the mixing angle is free from experimental constraints in the limit. 
Such a built-in cancellation mechanism is called a degenerate scalar scenario.
The extension of the scalar sector of the SM, as in the CxSM, has a possibility to realize the strong first-order electroweak phase transition (EWPT) required for the electroweak baryogenesis~\cite{Kuzmin:1985mm}. 
The EWPT in the degenerate scalar scenario of the CxSM has been studied in detail in Refs.~\cite{Cho:2021itv,Cho:2022our}. 

In general, some parameters, such as masses or couplings, are introduced in new physics models, and their values are adjusted by hand to be consistent with low-energy experimental data. The Multi-critical Point Principle (MPP) has been proposed  as a guiding principle for choosing the model parameters at a low-energy scale~\cite{Bennett:1993pj,Bennett:1996vy,Bennett:1996hx}. 
The MPP states that parameters of a theory take critical values so that multiple vacua have degenerate energy density. 
The application of the MPP by Froggatt and Nielsen to the effective potential of the SM Higgs was notable for predicting its mass accurately before the discovery of the Higgs boson~\cite{Froggatt:1995rt}. 
In Ref.~\cite{Froggatt:1995rt}, the MPP requires two degenerate vacua in the SM, one at the electroweak scale and another near the Planck scale. 
Since there is only one scalar field in the SM, it was necessary to discuss two vacua at different energy scales taking account of the radiative corrections to the Higgs potential. 
On the other hand, there are multiple vacua at a low-energy scale in new physics models with the extended scalar sector, such as the CxSM. 
The authors of Ref.~\cite{Kannike:2020qtw} claim that if the MPP is a fundamental principle, the MPP should be applied to all vacua, including ones dominated by the tree-level potential in models with the extended scalar sector. This is called the tree-level MPP. The model parameters are chosen to make multiple vacua degenerate at a low-energy scale by applying the tree-level MPP to models with the extended scalar sector. 

In this paper, we apply the tree-level MPP for the CxSM and study if the degenerate scalar scenario is realized. The application of the tree-level MPP to a variant of the CxSM has been discussed in Ref.~\cite{Kannike:2020qtw}. 
The authors analyze the scalar potential, which is called 
the pseudo Nambu-Goldstone DM model~\cite{Gross:2017dan}, and they pointed out that no degenerate vacua at low-energy scale exist in their model, i.e., the tree-level MPP does not allow the pseudo Nambu-Goldstone DM model. This model is, however, known to suffer from the domain-wall problem. We, therefore, adopt the most general and minimal scalar potential in the CxSM to avoid the domain-wall problem~\cite{Abe:2021nih}. We emphasize that this is the first study to apply the tree-level MPP to the degenerate scalar scenario in the CxSM. 

We show that the tree-level MPP allows both vacua with and without degenerate scalars in the CxSM. We then discuss the possibility of first-order EWPT in the model parameter space chosen by the tree-level MPP. 
It has been pointed out that the first-order EWPT in the CxSM is governed by the contribution from SU(2)$_L$ doublet-singlet mixing in the scalar potential at the tree-level~\cite{Cho:2021itv,Cho:2022our}. 
Therefore, since the dominant contribution to realizing the EWPT in the CxSM is given by the tree-level potential, requiring the tree-level MPP may break the relation of parameters which makes the first-order EWPT strong. 
However, we find that the 1-loop contribution would be enough to derive the first-order EWPT even after applying the tree-level MPP. 

The paper is organized as follows. In Sec.~\ref{sec:model}, we introduce the CxSM and describe the degenerate scalar scenario. Parameter space in the CxSM chosen by the tree-level MPP is analyzed in Sec.~\ref{sec:MPP}. The characteristic feature of the first-order EWPT in the CxSM and the consequence of imposing the tree-level MPP are also described. 
In Sec.~\ref{sec:result}, we study quantitatively whether first-order EWPT is feasible after imposing the tree-level MPP. Sec.~\ref{sec:Summary} is devoted to summarize our study.

\section{Model}\label{sec:model}

The CxSM consists of a complex scalar singlet $S$ in addition to the SM particles~\cite{Barger:2008jx}. 
In our study, we adopt the following scalar potential: 
\begin{align}
V_{0}(H, S)
 =
 \frac{m^{2}}{2} H^{\dagger} H+\frac{\lambda}{4}\left(H^{\dagger} H\right)^{2}+\frac{\delta_{2}}{2} H^{\dagger} H|S|^{2}+\frac{b_{2}}{2}|S|^{2}+\frac{d_{2}}{4}|S|^{4}+\left(a_{1} S+\frac{b_{1}}{4} S^{2}+\text{H.c.}\right), 
\label{tree}
\end{align} 
where a global U(1) symmetry for $S$ is softly broken by both $a_1$ and $b_1$. In the following, all the couplings in \eqref{tree} are assumed to be real. 
When the linear term of $S$ is absent, there is a $Z_2$ symmetry ($S\to -S$) in the scalar potential. Once the singlet $S$ develops the vacuum expectation value (VEV), the $Z_2$ symmetry is spontaneously broken, and it causes the domain-wall problem~\cite{Abe:2021nih}. 
We, therefore, add the linear term of $S$ to the scalar potential \eqref{tree} because it explicitly breaks the $Z_2$ symmetry, and the model does not suffer from the domain-wall problem. Besides $S$ and $S^2$ terms, although other operators dropped here could softly break the global U(1) symmetry, we adopt a minimal set of operators that close under renormalization~\cite{Barger:2008jx}. 

We parametrize the scalar fields as
\begin{align}
H &=\left(\begin{array}{c}
G^{+} \\
\frac{1}{\sqrt{2}}\left(v+h+i G^{0}\right)
\end{array}\right), \label{Hcomponent}\\
S &=\frac{1}{\sqrt{2}}\left(v_{S}+s+i \chi\right), 
\label{Scomponent}
\end{align}
where $v~(\simeq 246~\text{GeV})$ and $v_S$ represent the VEVs of $H$ and $S$, respectively. 
The Nambu-Goldstone bosons $G^+$ and $G^0$ are eaten by $W$ and $Z$ bosons, respectively, after the electroweak symmetry breaking. 
Since we assumed no complex parameters in \eqref{tree}, the scalar potential is invariant under CP-transformation ($S\to S^*$). Therefore, the real and imaginary parts of $S$ do not mix, and the stability of $\chi$ is guaranteed, making it a DM candidate.

The first-order derivatives of $V_0$ with respect to $h,s$ are
\begin{align}
\frac{1}{v}\left\langle\frac{\partial V_0}{\partial h}\right\rangle &=\frac{m^2}{2}+\frac{\lambda}{4} v^2+\frac{\delta_2}{4} v_S^2=0, \label{tadpole1}\\
\frac{1}{v_S}\left\langle\frac{\partial V_0}{\partial s}\right\rangle &=\frac{b_2}{2}+\frac{\delta_2}{4} v^2+\frac{d_2}{4} v_S^2+\frac{\sqrt{2} a_1}{v_S}+\frac{b_1}{2}=0,  \label{tadpole2}
\end{align}
where $\langle\cdots\rangle$ is defined as taking all fluctuation fields to zero. Note that our vacuum is such that $H$ and $S$ both have VEVs, i.e., $\qty(\langle H \rangle,~\langle S \rangle)=(v,~v_S)$. We also consider the case when only $S$ takes a VEV $\qty(\langle H \rangle,~\langle S \rangle)=(0,~v_S^{\prime})$ to discuss the tree-level MPP in Sec.~\ref{sec:MPP}. In this case, the derivative is expressed as
\begin{align}
\frac{1}{v_S^{\prime}}\left\langle\frac{\partial V_0}{\partial s}\right\rangle &=\frac{b_2}{2}+\frac{d_2}{4} v_S^{\prime 2}+\frac{\sqrt{2} a_1}{v_S^{\prime}}+\frac{b_1}{2}=0.  \label{tadpole3}
\end{align}
Note that since nonzero $v_S$ is enforced by $a_1 \neq 0$, the vacuum ($v,0$) does not appear in this model.

The mass matrix of the CP-even states ($h,s$) is expressed as
\begin{align}
\mathcal{M}_S^2=\left(\begin{array}{cc}
\lambda v^2 / 2 & \delta_2 v v_S / 2 \\
\delta_2 v v_S / 2 & \Lambda^2 
\end{array}\right),\quad\Lambda^2 \equiv \frac{d_2}{2} v_S^2-\sqrt{2}\frac{a_1}{v_S}. \label{MM}
\end{align}
The mass matrix \eqref{MM} is diagonalized by an orthogonal matrix $O(\alpha)$ as
\begin{align}
O(\alpha)^\top \mathcal{M}_S^2 O(\alpha)=\left(\begin{array}{cc}
m_{h_1}^2 & 0 \\
0 & m_{h_2}^2
\end{array}\right), \quad O(\alpha)=\left(\begin{array}{cc}
\cos \alpha & -\sin \alpha \\
\sin \alpha & \cos \alpha
\end{array}\right), 
\label{masseigenstate}
\end{align}
where $\alpha$ is a mixing angle and a symbol $\top$ denotes the transpose of the matrix. 
The mass eigenstates ($h_1,~h_2$) are given through the mixing angle $\alpha$ as
\begin{align}
\left(\begin{array}{l}
h \\
s
\end{array}\right)=\left(\begin{array}{cc}
\cos \alpha & \sin \alpha \\
-\sin \alpha & \cos \alpha
\end{array}\right)\left(\begin{array}{l}
h_1 \\
h_2
\end{array}\right).
\end{align}
We emphasize that $\alpha \to 0$ corresponds to the SM-like limit $(h_1 \to h,~h_2 \to s)$. The mass eigenvalues are expressed as
\begin{align}
m_{h_1, h_2}^2 &=\frac{1}{2}\left(\frac{\lambda}{2} v^2+\Lambda^2 \mp \frac{\frac{\lambda}{2} v^2-\Lambda^2}{\cos 2 \alpha}\right) \\
&=\frac{1}{2}\left(\frac{\lambda}{2} v^2+\Lambda^2 \mp \sqrt{\left(\frac{\lambda}{2} v^2-\Lambda^2\right)^2+4\left(\frac{\delta_2}{2} v v_S\right)^2}\right),
\end{align}
We fix $h_1$ as the Higgs boson observed in the LHC experiments, i.e., $m_{h_1}=125$~GeV. 
The mass of CP-odd state $\chi$ is given by the soft breaking terms $a_1$ and $b_1$ as 
\begin{align}
m_\chi^2
 &=
 \frac{b_2}{2}-\frac{b_1}{2}+\frac{\delta_2}{4} v^2+\frac{d_2}{4} v_S^2\nonumber \\
 &=
 -\frac{\sqrt{2} a_1}{v_S}-b_1. \label{DMmass}
\end{align}

For later convenience, we mention the relationship between input and output parameters. 
In the following study, we adopt $\left\{v, v_S, m_{h_1}, m_{h_2}, \alpha, m_\chi, a_1\right\}$ as inputs while 
the Lagrangian parameters $\left\{m^2, b_2, \lambda, d_2, \delta_2, b_1\right\}$ can be expressed as functions of inputs. 
Among the Lagrangian parameters, $m^2$ and $b_2$ are eliminated from the tadpole conditions (\ref{tadpole1}) and (\ref{tadpole2}),
\begin{align}
m^2 &=-\frac{\lambda}{2} v^2-\frac{\delta_2}{2}v_S^2  \label{msquare}\\
b_2 &=-\frac{\delta_2}{2} v^2-\frac{d_2}{2}v_S^2-\sqrt{2}\frac{a_1}{v_S}-b_1. \label{b2}
\end{align}
The remaining four parameters in the six Lagrangian parameters are given as 
\begin{align}
\lambda&=\frac{2}{v^2}\left(m_{h_1}^2\cos^2\alpha+m_{h_2}^2\sin^2\alpha\right),\label{lambda}\\
\delta_2&=\frac{1}{v v_S}\left(m_{h_1}^2-m_{h_2}^2\right)\sin{2\alpha}\label{del2},\\
d_2&=2\left(\frac{m_{h_1}}{v_S}\right)^2\sin^2\alpha+2\left(\frac{m_{h_2}}{v_S}\right)^2\cos^2\alpha+2\sqrt{2}\frac{a_1}{v_S^3}\label{d2},\\
b_1 &=-m_{\chi}^2-\frac{\sqrt{2}}{v_S}a_1. 
\end{align}

Theoretical constraints on the quartic couplings in the scalar potential are summarized as follows. 
A requirement on the scalar potential that is bounded from below is given by\footnote{For $\delta_2<0$, $\lambda d_2>\delta_2^2$ is also needed. In this paper,  we assume that $\delta_2$ is positive.}
\begin{align}
\lambda>0,\quad d_2 > 0.
\label{bfb}
\end{align} 
The couplings $\lambda$ and $d_2$ should also satisfy the following condition from the perturbative unitarity~\cite{Abe:2021nih}
\begin{align}
\lambda < \frac{16\pi}{3},\quad d_2< \frac{16\pi}{3}.
\end{align}
In addition, the stability condition of the tree-level potential requires~\cite{Barger:2008jx}
\begin{align}
\lambda\left(d_2-\frac{2 \sqrt{2} a_1}{v_S^3}\right)>\delta_2^2.
\label{stability}
\end{align}
Recent DM direct detection experiments have provided upper limits on the spin-independent cross section of DM scattering off nucleons ($\sigma_{\text{SI}}$)~\cite{XENON:2020kmp,LZ:2022ufs}, and those results give a severe constraint on a certain class of DM models. 
In the CxSM, the DM-nucleon scattering process is mediated by two scalars ($h_1$ and $h_2$). In the following, we briefly review the degenerate scalar scenario~\cite{Abe:2021nih} where the DM-quark scattering is suppressed when masses of $h_1$ and $h_2$ are degenerate. 
The interaction Lagrangian of DM $\chi$ to the CP-even scalars $h_1$, $h_2$ is given by 
\begin{align}
\mathcal{L}_S 
 &=
 -\frac{m_{h_1}^2+\frac{\sqrt{2} a_1}{v_S}}{2 v_S} \sin \alpha~ h_1 \chi^2+\frac{m_{h_2}^2+\frac{\sqrt{2} a_1}{v_S}}{2 v_S} \cos \alpha~ h_2 \chi^2, 
\label{cch} 
\end{align}
while that of a quark $q$ to $h_1$ or $h_2$ is given by
\begin{align}
\mathcal{L}_Y&=\frac{m_q}{v} \bar{q} q \left(h_1 \cos \alpha-h_2 \sin \alpha\right) \label{hff}, 
\end{align}
where $m_q$ denotes a mass of the quark $q$. 
Then the scattering amplitudes $\mathcal{M}_{h_1}$ and $\mathcal{M}_{h_2}$ 
mediated by $h_1$ and $h_2$, respectively, are  
\begin{align}
&i \mathcal{M}_{h_1}=-i \frac{m_q}{v v_S} \frac{m_{h_1}^2+\frac{\sqrt{2} a_1}{v S}}{t-m_{h_1}^2} \sin \alpha \cos \alpha~\bar{u}\left(p_3\right) u\left(p_1\right), \label{amph1}\\
&i \mathcal{M}_{h_2}=+i \frac{m_q}{v v_S} \frac{m_{h_2}^2+\frac{\sqrt{2} a_1}{v_S}}{t-m_{h_2}^2} \sin \alpha \cos \alpha~\bar{u}\left(p_3\right) u\left(p_1\right), \label{amph2}
\end{align}
where $u(p_1)$ ($\bar{u}(p_3)$) is incoming (outgoing) quark-spinor with a momentum $p_1$ ($p_3$), and $t$ is defined as $t \equiv (p_1 - p_3)^2$. 
Since the momentum transfer $t$ in the direct detection experiments is small, 
the amplitude with $t\to 0$ becomes 
\begin{align}
i\left(\mathcal{M}_{h_1}+\mathcal{M}_{h_2}\right) &\simeq i \frac{m_q}{v v_S} \sin \alpha \cos \alpha~\bar{u}\left(p_3\right) u\left(p_1\right) \frac{\sqrt{2} a_1}{v_S}\left(\frac{1}{m_{h_1}^2}-\frac{1}{m_{h_2}^2}\right).
\label{DMamplitude}
\end{align}
We find that the sum of two amplitudes is highly suppressed when $m_{h_1} \simeq m_{h_2}$ because of the sign difference between \eqref{amph1} and \eqref{amph2}, which is due to the orthogonality of the mixing matrix $O(\alpha)$~\eqref{masseigenstate}. 
We note that, as pointed out in~Ref.~\cite{Gross:2017dan} , the two scattering amplitudes could also cancel each other when $a_1 \to 0$ without requiring the mass degeneracy of $h_1$ and $h_2$, which is known as the pseudo Nambu-Goldstone DM model. In such a case, as mentioned earlier, the scalar potential has the $Z_2$ symmetry, and it suffers from the domain-wall problem. 

The possibility of searching for the degenerate scalar scenario at collider experiments has been studied in Ref.~\cite{Abe:2021nih}. 
The authors stated that, although the mass difference $\left|m_{h_1}-m_{h_2}\right| \lesssim 3~\mathrm{GeV}$ is not ruled out from the LHC experiments~\cite{CMS:2014afl}, $\left|m_{h_1}-m_{h_2}\right| \lesssim 1~\mathrm{GeV}$ may be testable at the future $e^+ e^-$ linear collider. 

We mention that the degenerate scalar scenario is compatible with the Higgs search experiments at the LHC. 
As shown in Eq.~(\ref{hff}), the couplings between $h_1~(h_2)$ and the SM particles are those with the SM Higgs boson multiplied by $\cos{\alpha}~(-\sin{\alpha})$. For example, decay rates from $h_1$ and $h_2$ to the SM particle $X$ is expressed as follows; 
\begin{align}
\Gamma_{h_1\to XX}&=\cos^2{\alpha}~\Gamma_{h\to XX}^{\mathrm{SM}}(m_{h_1}), \label{partialdecaywidth}\\
\Gamma_{h_2\to XX}&=\sin^2{\alpha}~\Gamma_{h\to XX}^{\mathrm{SM}}(m_{h_2}),
\end{align}
where $\Gamma_{h\to XX}^{\mathrm{SM}}(m_{h_{1(2)}})$ is the Higgs partial decay width in the SM as a function of $m_{h_{1(2)}}$. Experimentally, when two scalars are degenerate ($m_{h_1} \simeq m_{h_2}$), it is hard to distinguish the production and decay processes of $h_2$ from those of $h_1$ so that the sum of two processes by $h_1$ and $h_2$ is to be observed, i.e., 
\begin{align}
\Gamma_{h_1\to XX}+\Gamma_{h_2\to XX} \simeq \Gamma_{h\to XX}^{\mathrm{SM}}(m_h), 
\end{align}
holds for any $\alpha$. 
Therefore, the signal strength of Higgs bosons in the CxSM is identical to that in the SM in the degenerate limit of two scalars.

\section{Tree-level MPP and EWPT in the CxSM}\label{sec:MPP}

We apply the tree-level MPP to the CxSM whose the tree-level scalar potential is given by Eq.~(\ref{tree}). 
Note that since $v_S$ is nonzero due to $a_1\neq0$, there are two possible vacua; the electroweak vacuum~$(v,v_S)$ and the singlet vacuum~$(0,v_S^{\prime})$. We consider the case where these two vacua are degenerate. The difference between the energy densities at these vacua is expressed as
\begin{align}
\Delta V_0
&\equiv V_0(v,v_S)-V_0(0,v_S^{\prime}) \nonumber \\
&=\frac{m^2}{8}v^2+\frac{3\sqrt{2} a_1}{4}(v_S-v_S^{\prime})+\frac{b_1+b_2}{8}(v_S^2-v_S^{\prime 2}). \label{potentialdifference}
\end{align} 
When $a_1=0$, $\Delta V_0$ is evaluated as 
\begin{align}
\Delta V_0
 &=\frac{m^2}{8}v^2+\frac{b_1+b_2}{8}(v_S^2-v_S^{\prime 2})
\nonumber \\
&\propto -\frac{1}{\lambda d_2-\delta_2^2} \frac{1}{d_2} \times\left[\delta_2\left(b_2+b_1\right)-d_2 b_2\right]^2<0. 
\label{potentialdifferenceZ2}
\end{align}
Since the denominator in \eqref{potentialdifferenceZ2} is positive due to \eqref{stability} with $a_1=0$, $\Delta V_0$ is always negative when the global U(1) symmetry is softly broken via only $S^2$ term, i.e., there are no degenerate vacua in this case~\cite{Kannike:2020qtw}. This result is altered by introducing the linear term of $S$ to the scalar potential, as shown in the following.  

To begin, we qualitatively discuss the model parameters of the CxSM required by the tree-level MPP. 
It is easy to see that one needs $v_S \neq v_S^{\prime}$ to degenerate two vacua because 
the first term of the r.h.s. in (\ref{potentialdifference}) is nonzero and negative (see, Eq.~(\ref{msquare})). 
Two VEVs, $v_S$ and $v_S^{\prime}$, are derived by tadpole conditions (\ref{tadpole2}) and (\ref{tadpole3}), respectively, the difference being the presence $\delta_2$ (\ref{del2}) which represents the doublet-singlet mixing. 
Therefore, $\delta_2$ and $v_S-v_S^{\prime}$ should be sizable to cancel with the first term in (\ref{potentialdifference}), to achieve $\Delta V_0=0$. 

From Eq.~(\ref{del2}), we find that $v_S$ should be small to make $\delta_2$ sizable whether we adopt the degenerate scalar scenario. In the degenerate scalar scenario, the  mass difference ($m_{h_1}^2 - m_{h_2}^2$ in \eqref{del2} ) is small,  so a large $\delta_2$ is possible by requiring small $v_S$. 
In non-degenerate case ($m_{h_1} \neq m_{h_2}$), 
the couplings of $h_1$ with $m_{h_1}=125~\mathrm{GeV}$ to the SM particles given by multiplying $\cos\alpha$ to the SM Higgs couplings. Taking account of the Higgs search experiments at the LHC, $\cos\alpha$ should be close to 1, i.e., $\sin 2\alpha$ in \eqref{del2} must be small. 
Therefore $v_S$ in the denominator in \eqref{del2} should also be small to make $\delta_2$ sizable. 

We quantitatively discuss the parameter space in the CxSM with degenerate scalars chosen by applying the tree-level MPP. A case where two scalars are not degenerate is also studied for comparison. 
In the following numerical study,  we set $m_\chi=62.5$ GeV as a reference. We note that the DM mass makes almost no contribution to $\Delta V_0$ since only $b_1$ is an output parameter of  DM mass $m_\chi$~(\ref{DMmass}) while the potential difference $\Delta V_0$ depends on $b_1+b_2$ which does not include $b_1$ term as found in Eq.~(\ref{b2}).


\begin{figure}
\center
\includegraphics[width=8.1cm]{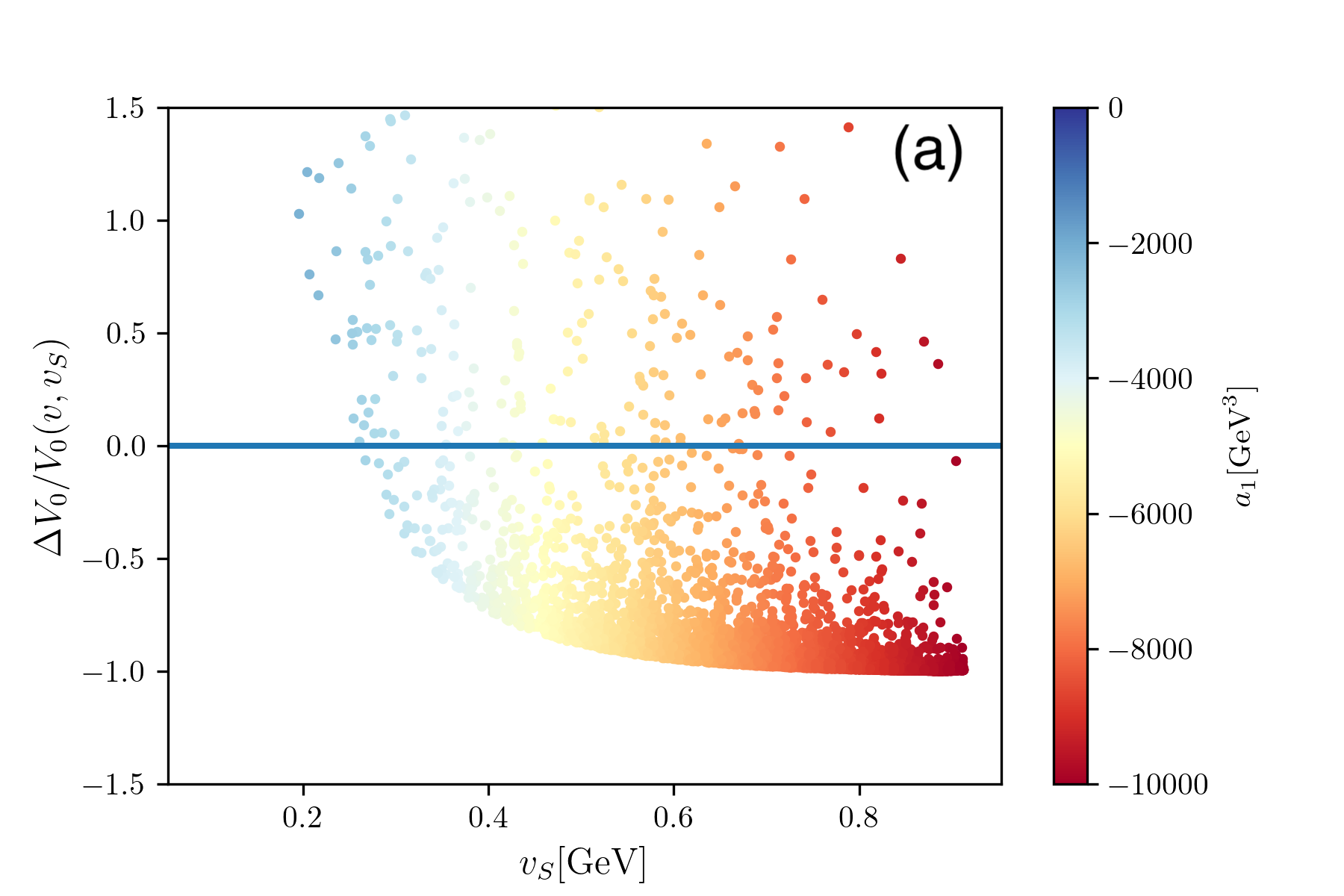}
\includegraphics[width=8.1cm]{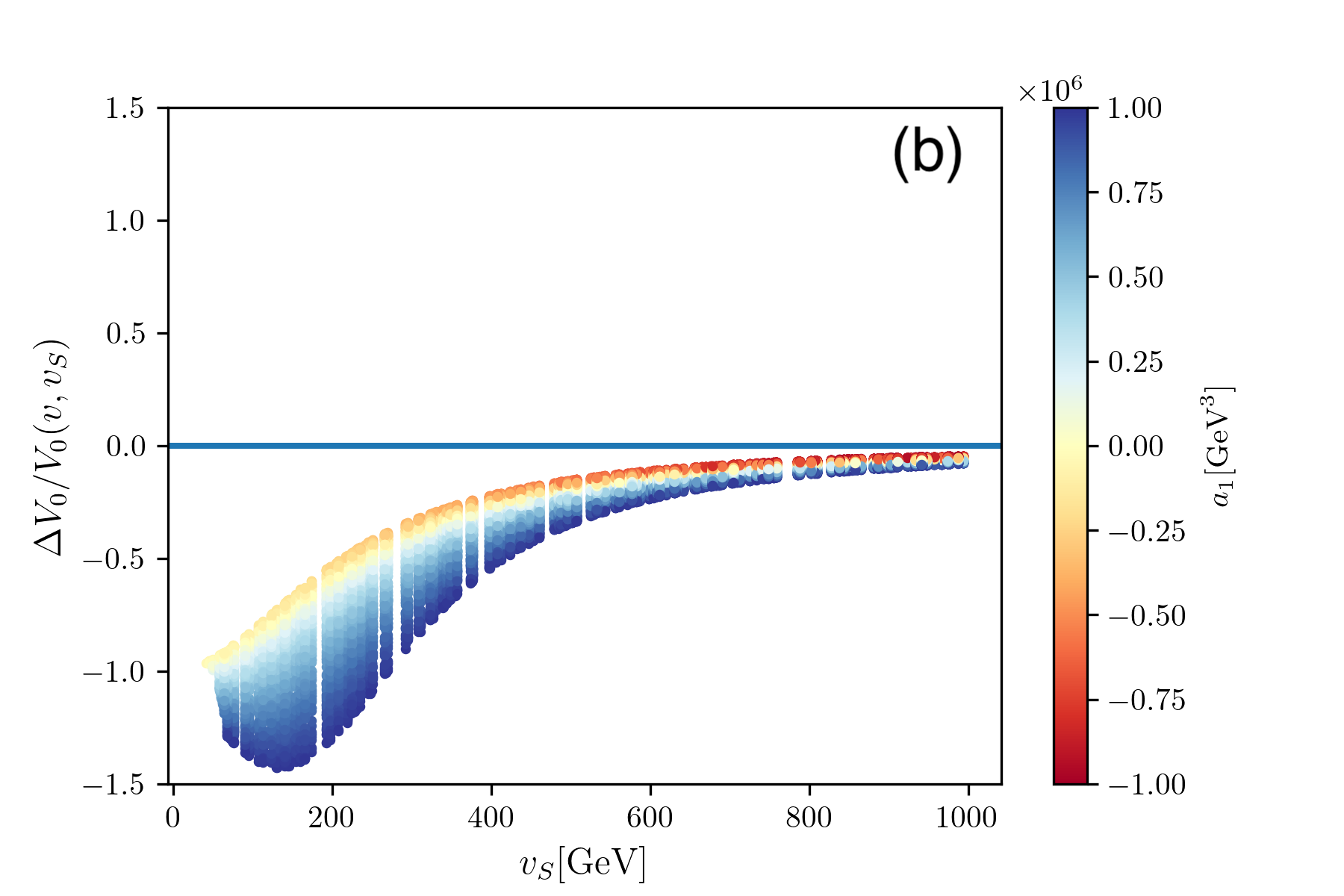}
\caption{The potential difference for $v_S = 0.1 - 1~\mathrm{GeV}$~(a) and $v_S = 100-1000~\mathrm{GeV}$~(b) in the degenerate scalar scenario. The color bar represents the change in $a_1$. }
\label{degenerateMPP}
\end{figure}

First, we consider the degenerate scalar scenario. The second Higgs mass $m_{h_2}$ is set to $124~\mathrm{GeV}$. As mentioned in Section \ref{sec:model}, when the masses of two Higgs bosons are degenerate, there are no constraints on the mixing angle $\alpha$. Here we adopt the mixing angle $\alpha=\pi/4$ as an example. Fig.~\ref{degenerateMPP} shows the potential difference scaled by the energy density of the electroweak vacuum $\Delta V_0/V_0(v,v_S)$ as a function of $v_S$. The range of $v_S$ is $0.1 - 1~\mathrm{GeV}$ (Fig.~\ref{degenerateMPP}(a)) and $100 -1000~\mathrm{GeV}$ (Fig.~\ref{degenerateMPP}(b)).  Taking account of the theoretical constraints on the scalar potential summarized in Eqs.~\eqref{bfb}-\eqref{stability}, we found no allowed parameter space for $v_S= 10 - 100~\mathrm{GeV}$. Therefore we drop the range $v_S= 10 - 100~\mathrm{GeV}$ from the figure. 
The color bar represents the $a_1$ dependence. When $v_S$ is small (Fig.~\ref{degenerateMPP}(a)), $\Delta V_0/V_0(v,v_S)$ can be either positive or negative. On the other hand, when $v_S$ is large (Fig.~\ref{degenerateMPP}(b)), $\Delta V_0/V_0(v,v_S)$ approaches zero but never reaches zero. Therefore, degenerate vacua are realized only when $v_S=\mathcal{O}(0.1)$.

\begin{figure}
\center
\includegraphics[width=8.1cm]{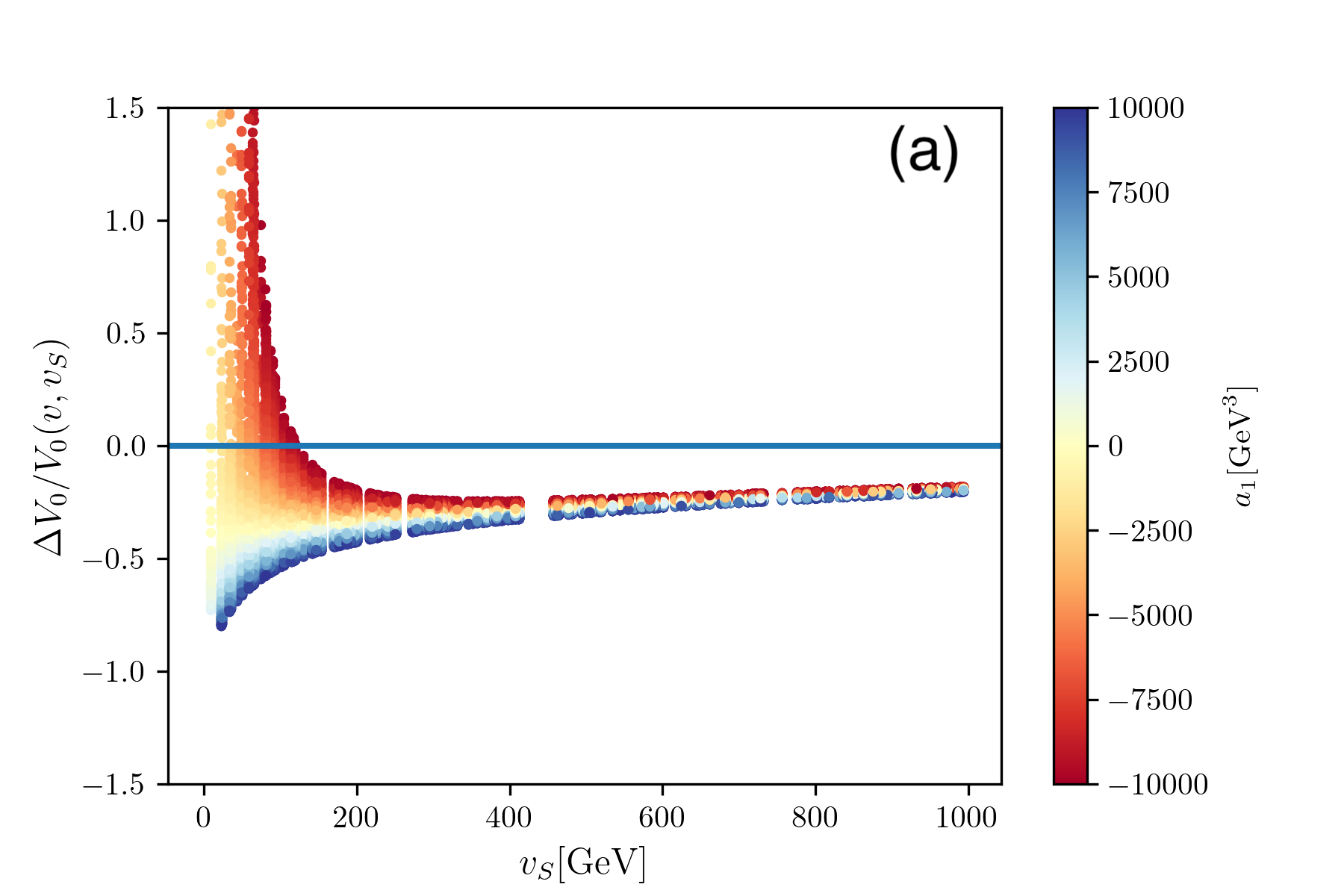}
\includegraphics[width=8.1cm]{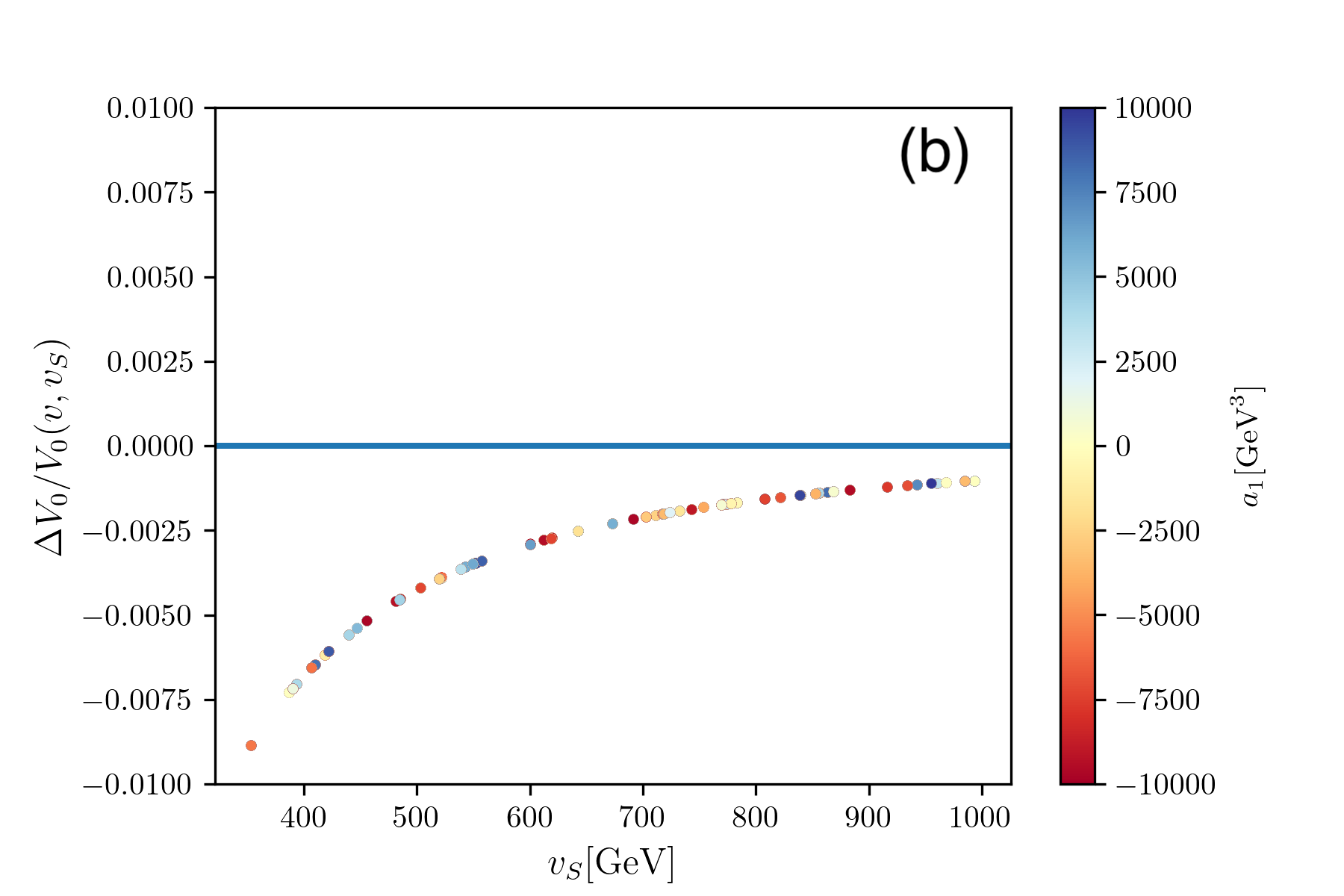}
\caption{The potential difference for $v_S$ when $m_{h_2}$= 10 GeV~(a) and $m_{h_2}$ =1000 GeV~(b) in the non-degenerate Higgs case. The color bar represents the change in $a_1$.}
\label{nondegenerateMPP}
\end{figure}

Next, we check for the possible existence of multi-critical points when the masses of two scalars are far apart. If the mass difference of two scalars is large, the couplings of the second scalar $h_2$ to the SM particles are proportional to $\sin\alpha$, and no signal of an additional scalar at the LHC requires the mixing angle $\alpha$ to be small. So we set $\alpha=0.1$ in our study\footnote{In Sec.\ref{sec:result}, we set benchmark points that satisfy the LHC constraints.}. Fig.~\ref{nondegenerateMPP} shows $\Delta V_0/V_0(v,v_S)$ as a function of $v_S$ for $m_{h_2}$ =10 GeV~(Fig.~\ref{nondegenerateMPP}(a)) and $m_{h_2}$ =1000 GeV~(Fig.~\ref{nondegenerateMPP}(b)). The color bar represents the value of $a_1$. When $m_{h_2}$ = 10 GeV, for $v_S$ = 0.1 - 1 GeV, the theoretical constraints on parameters in the scalar potential~\eqref{bfb}-\eqref{stability}
are not satisfied. 
The degenerate vacua $\Delta V_0/V_0(v,v_S)=0$ 
are seen in small $v_S$ as in the degenerate scalar scenario. However, possible $v_S$ regions are relaxed. On the other hand, When $m_{h_2}$ = 1000 GeV, the theoretical constraints are not satisfied for small $v_S$, thus, inevitably, large $v_S$ is required, and the tree-level MPP is not viable.


We now discuss the first-order phase transition in the CxSM and mention the compatibility of EWPT and the tree-level MPP. The 1-loop effective potential, which is used to study the phase transition, can be written in this way;
\begin{align}
V_{\text{eff}}\left(\varphi, \varphi_S ; T\right)&=V_{0}(\varphi, \varphi_S)+V_{1}(\varphi, \varphi_S;T), \nonumber \\
&=V_{0}(\varphi, \varphi_S)+\sum_i n_i\left[V_{\mathrm{CW}}\left(\bar{m}_i^2\right)+\frac{T^4}{2 \pi^2} I_{B, F}\left(\frac{\bar{m}_i^2}{T^2}\right)\right], \label{eff}
\end{align}
where $T$ represents temperature and $\varphi$ and $\varphi_S$ are the constant classical background fields of $H$ and $S$, respectively. $V_{0}$ represents the tree-level potential. The indices $i$ express $h_{1,2}$, $\chi$, $W$, $Z$, $t$ and $b$. The degrees of freedom of each particle $n_i$ are $n_{h_{1,2},\chi}=1, n_W=6, n_Z=3, \text { and } n_t=n_b=-12$. The first term in parentheses in the second row in (\ref{eff}) is the $\overline{\text{MS}}$-defined effective potential at zero temperature, and the second one is the finite temperature effective potential. They are given by~\cite{Weinberg:1973am,Jackiw:1974cv,Dolan:1973qd}
\begin{align}
V_{\mathrm{CW}}\left(\bar{m}_i^2\right) &=\frac{\bar{m}_i^4}{64 \pi^2}\left(\ln \frac{\bar{m}_i^2}{\bar{\mu}^2}-c_i\right), \\
I_{B, F}\left(a^2\right) &=\int_0^{\infty} d x x^2 \ln \left(1 \mp e^{-\sqrt{x^2+a^2}}\right) \label{finite},
\end{align}
where $\bar{m_i}$ is the field-dependent masses of each particle $i$. $c_i=3/2$ for scalars and fermions and $c_i=5/6$ for gauge bosons.  $I_B$ with the minus sign represents the boson contribution, while $I_F$ with the plus sign represents the fermion one.  $\bar{\mu}$ is a renormalization scale. The potential $V_{\mathrm{CW}}$ is renormalized so that the scalar mass and the tad-pole conditions take the same relationship with those at the tree-level. 

To achieve EWBG, EWPT must be strong first-order, which requires the following condition~\cite{Arnold:1987mh,Bochkarev:1987wf,Funakubo:2009eg};
\begin{align}
\frac{v_C}{T_C} \gtrsim 1,\label{decoupling}
\end{align}
where $T_C$ is the critical temperature defined as one when the effective potential has two degenerate minima, and $v_C$ is the Higgs VEV at $T=T_C$. The lower bound in Eq. (\ref{decoupling}) depends on the structure of the saddle point of classical solutions, called sphaleron. 

Now we discuss the compatibility of the EWPT and the tree-level MPP in the CxSM. 
As is already mentioned, the critical temperature $T_C$ is defined as the temperature at which the effective potential has two degenerate vacua;
\begin{align}
V_{\mathrm{eff}}\left(v_C, v_{SC} ;T_C\right)=V_{\mathrm{eff}}\left(0, v_{SC}^{\prime} ;T_C\right) 
&\to 
 \nonumber\\
V_{0}\left(v_C, v_{SC}\right)+V_{1}\left(v_C, v_{SC} ;T_C\right)&=V_{0}\left(0, v_{SC}^{\prime}\right)+V_{1}\left(0, v_{SC}^{\prime} ;T_C\right),
\label{TCdefinition}
\end{align}
with
\begin{align}
v_C=\lim _{T \to T_C} v(T), \quad v_{S C}=\lim _{T \to T_C} v_S(T), \quad v_{S C}^{\prime}=\lim _{T \to T_C} v_S^{\prime}(T).
\end{align}
On the other hand, the vacuum degeneracy required by the tree-level MPP is
\begin{align}
V_{0}\left(v, v_{S}\right)=V_{0}\left(0, v_{S}^{\prime}\right), \label{MPPdefinition}
\end{align}
with
\begin{align}
v=\lim _{T \to 0} v(T), \quad v_{S}=\lim _{T \to 0} v_S(T), \quad v_{S}^{\prime}=\lim _{T \to 0} v_S^{\prime}(T).
\label{vevt}
\end{align}

It is known that, in general, the first-order EWPT is induced by 
the bosonic contributions of the finite temperature potential at the 1-loop level. 
However, in a certain class of models with the extended scalar sector, such as the CxSM, the tree-level structure of the scalar potential is rather crucial to realize the EWPT (see, Ref.~\cite{Cho:2021itv,Cho:2022our} for details)\footnote{For classification of first-order EWPT, refer, e.g., Ref.~\cite{Chung:2012vg}.}. 
Now, Eqs.~\eqref{TCdefinition}-\eqref{vevt} tell us that, 
except for the 1-loop effective potential $V_1$, the definition of $T_C$~(\ref{TCdefinition}) is the same with Eq.~(\ref{MPPdefinition}) derived from the tree-level MPP. 
In other words, when the tree-level MPP is valid, EWPT occurs at zero temperature, which conflicts with the condition (\ref{decoupling}).  
Therefore, by imposing the tree-level MPP to the CxSM, EWPT from the tree-level potential cannot be defined. On the other hand, there are subleading contributions to the EWPT at the 1-loop level from additional bosons in the CxSM. 
Thus, in the next section, we evaluate if the subleading contributions are enough for the strong first-order EWPT at some benchmark points where the tree-level MPP is valid.


\section{Numerical results}\label{sec:result}

To discuss the strong first-order EWPT quantitatively, we focus on two benchmark points from the model parameter space chosen by the tree-level MPP (see, Table~\ref{tab:BP}). 
We select these points such that one (BP1) corresponds to the parameter set where two scalars are degenerate ($m_{h_1} \simeq m_{h_2}$) while another (BP2) is not  ($m_{h_1} \neq m_{h_2}$). 
The coefficient $a_1$ is set to make $\Delta V_0$~\eqref{potentialdifference} to zero.
\begin{table}[t]
\center
\begin{tabular}{|c|c|c|c|c|c|c|c|c|}
\hline
Inputs & $v$ [GeV] & $v_S$ [GeV] & $m_{h_1}$ [GeV] & $m_{h_2}$ [GeV] & $\alpha$ [rad]  & $m_\chi$ [GeV] & $a_1$ [GeV$^3$]\\ \hline
BP1 & 246.22 & 0.6  & 125.0 & 124.0  & $\pi/4$ & 62.5 & $-6576.2385$ \\ \hline
BP2 & 246.22 & 10.0  & 125.0 & 10.0  & $0.001$ & 62.5 & $-707.1913$ \\ \hline\hline
Outputs & $m^2$ & $b_2$ [GeV$^2$] & $b_1$ [GeV$^2$] & $\lambda$ & $\delta_2$ & $d_2$ & $a_1$ [GeV$^3$]   \\ \hline
BP1 & $-(124.5)^2$ & $-(178.0)^2$ & $(107.7)^2$ & 0.511 &1.69 & 0.87 &  $-6576.2385$  \\ \hline
BP2 & $-(125.0)^2$ & $(60.20)^2$ & $-(61.69)^2$ & 0.515 & 0.013& $7.1\times10^{-5}$ &  $-707.1913$  \\ \hline
\end{tabular}
\caption{Input and output parameters in the two benchmark points. }
\label{tab:BP}
\end{table}
First, we show that two benchmark points satisfy constraints on the DM relic density $\Omega_{\mathrm{DM}}h^2$~\cite{Planck:2018vyg};
\begin{align}
\Omega_{\mathrm{DM}} h^2=0.1200 \pm 0.0012, \label{relic}
\end{align}
 and the spin-independent scattering cross section with the nucleons $\sigma_{\mathrm{SI}}$ by the LUX-ZEPLIN (LZ) experiment~\cite{LZ:2022ufs}. 
In the following numerical study, we use a public code \texttt{micrOMEGAs}~\cite{Belanger:2020gnr}. The relic density of DM $\chi$ in the CxSM, $\Omega_\chi h^2$, should not exceed the observed value~\eqref{relic} and, as will be shown later, can explain only a portion of $\Omega_{\mathrm{DM}}h^2$. 
Therefore, for $\Omega_\chi<\Omega_{\mathrm{DM}}$, we scale $\sigma_{\mathrm{SI}}$ as
\begin{align}
\tilde{\sigma}_{\mathrm{SI}}=\left(\frac{\Omega_\chi}{\Omega_{\mathrm{DM}}}\right) \sigma_{\mathrm{SI}},
\end{align}
and this should satisfy the bound from the LZ experiment. 

\begin{figure}
\center
\includegraphics[width=8.1cm]{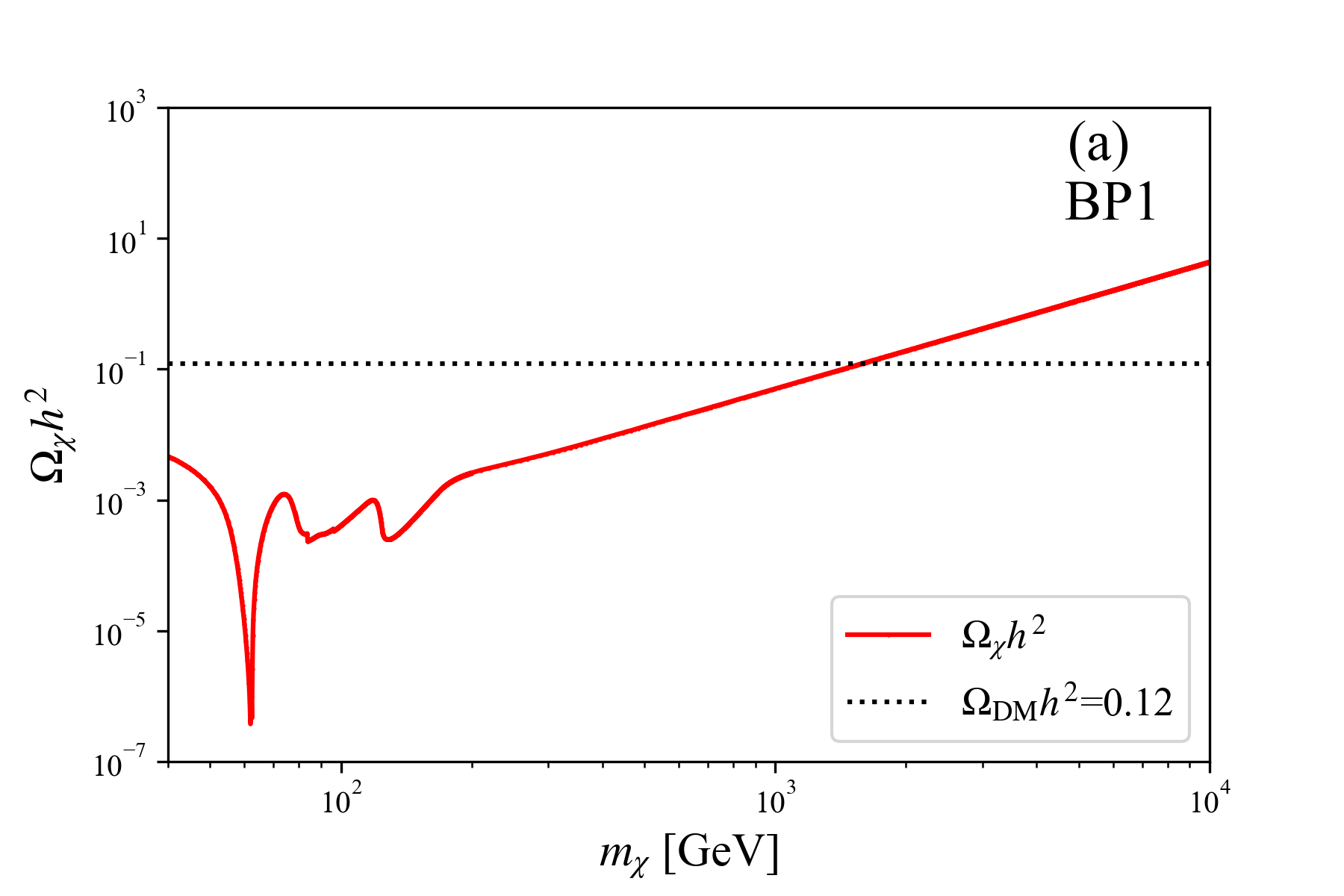}
\includegraphics[width=8.1cm]{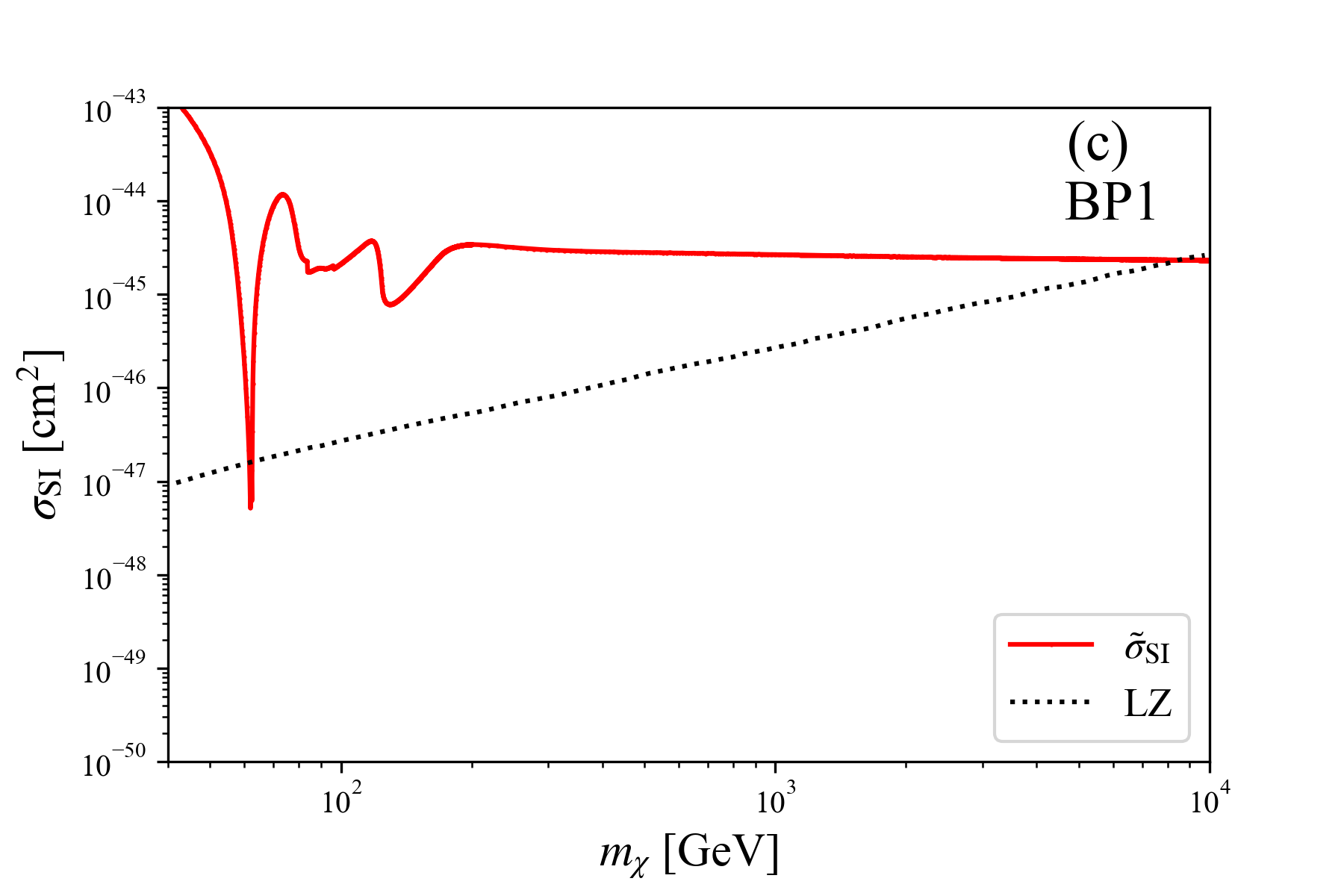}\\
\includegraphics[width=8.1cm]{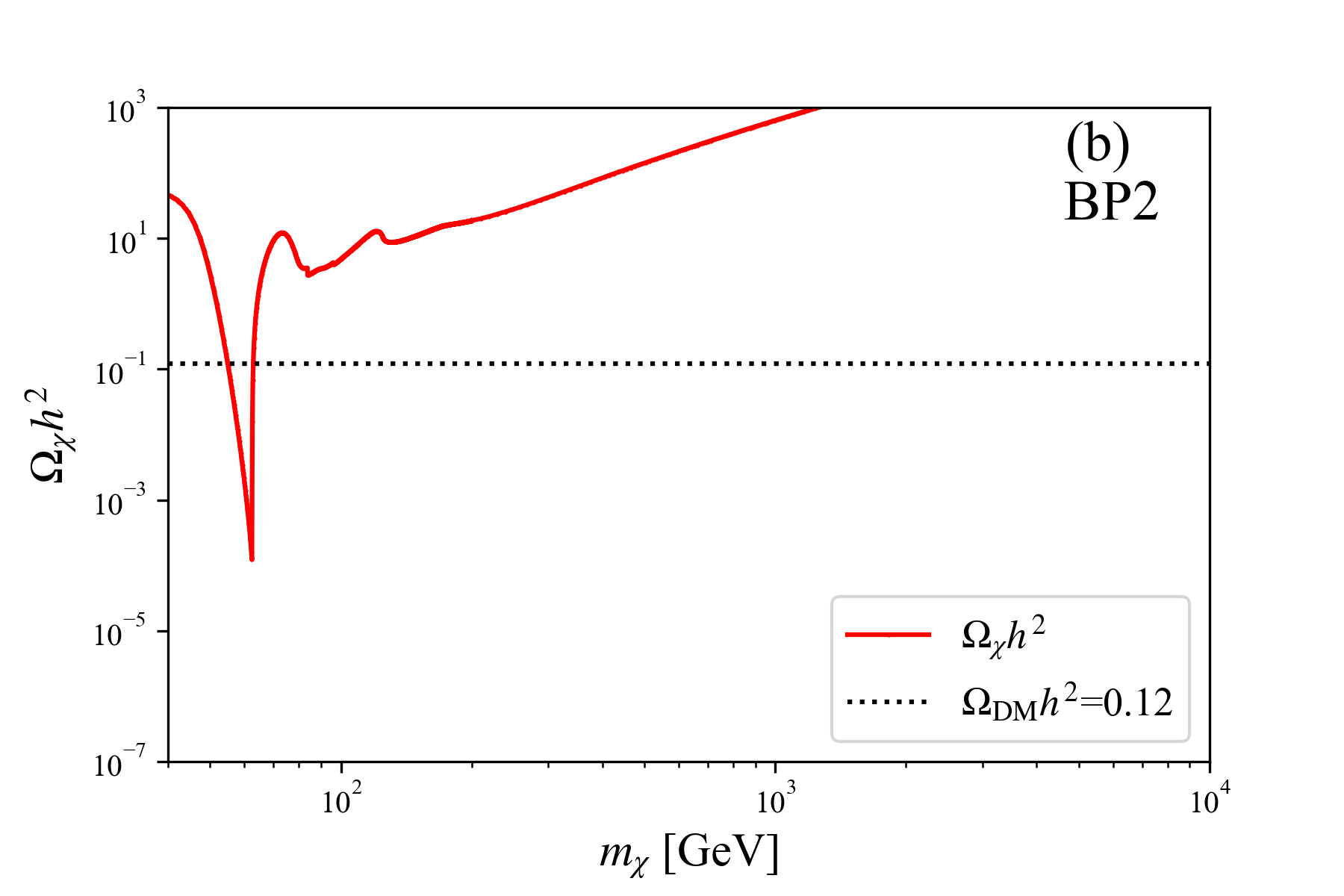}
\includegraphics[width=8.1cm]{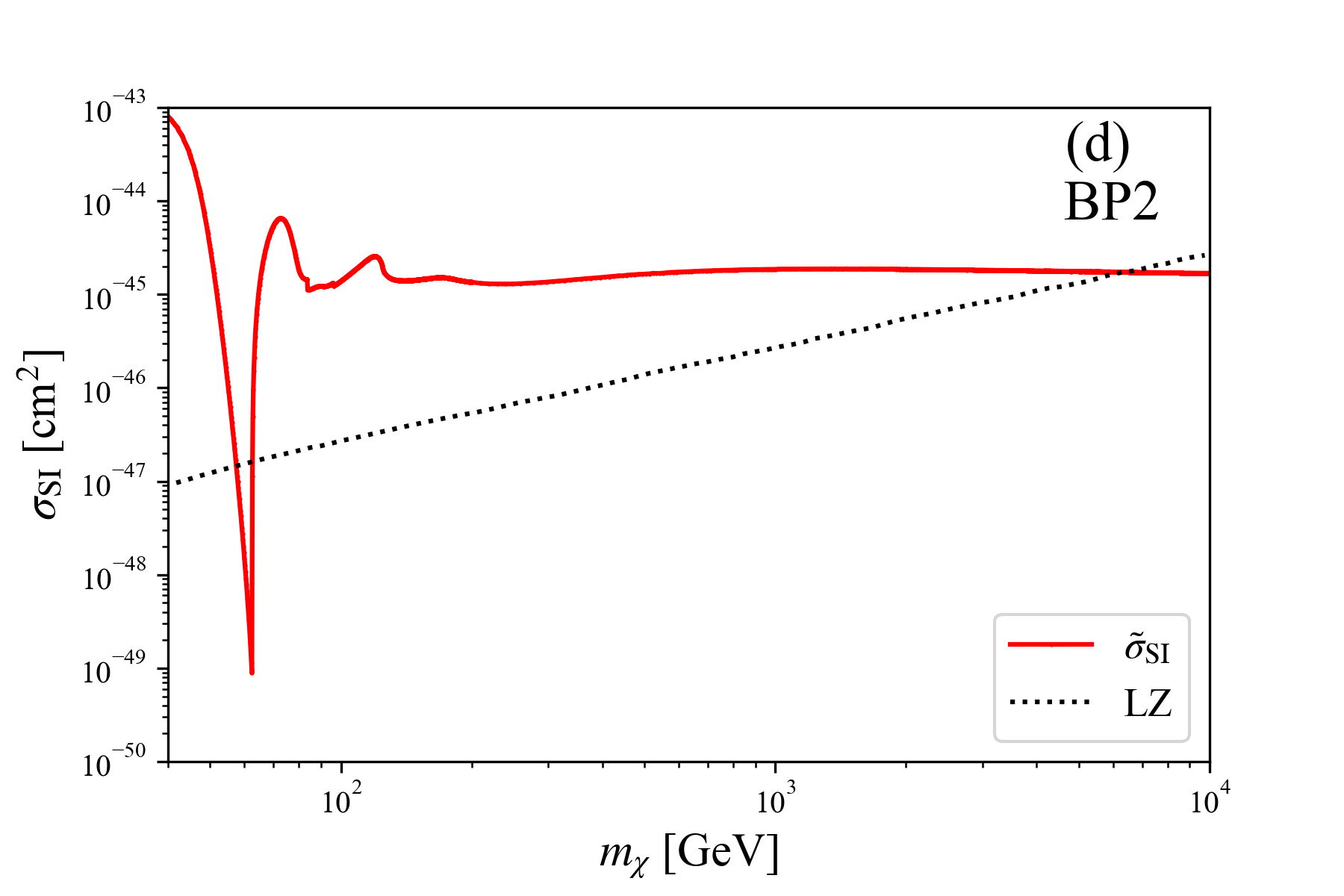}
\caption{DM relic density $\Omega_{\chi}h^2$~(a),(b)
and scaled spin-independent scattering cross section with the nucleons $\tilde{\sigma}_{\mathrm{SI}}$~(c),(d) are shown as a function of the DM mass $m_\chi$. The upper panels~(a),(c) correspond to BP1, and the lower ones to BP2~(b),(d). The dotted lines in the left panels~(a),(c) show the observed relic density, and the dotted curves in the right panels~(b),(d) show the LZ experimental results.
}
\label{BP1BP2DM}
\end{figure}
Two left figures in Fig.~\ref{BP1BP2DM} show $\Omega_{\chi}h^2$ as a function of $m_\chi$ at the benchmark points BP1 (a) and BP2 (b). 
In the figures,  the dotted horizontal line denotes a central value of the observed DM relic density $\Omega_{\mathrm{DM}} h^2$~\eqref{relic}. 
In Fig.~\ref{BP1BP2DM}~(a) (BP1), we find that $\Omega_{\chi}h^2$ is smaller than $\Omega_{\mathrm{DM}} h^2=0.12$ up to around $m_\chi$= 1500 GeV. On the other hand, in Fig.~\ref{BP1BP2DM}~(b) (BP2), 
$\Omega_\chi h^2$ at only around $m_\chi$= 62.5 GeV is allowed compared to $\Omega_{\mathrm{DM}} h^2=0.12$. 
The dip around $m _\chi$= 62.5 GeV~(= $m_{h_1}/2$) seen at both benchmark points reflects resonance enhancement due to $s$-channel DM annihilation processes. 

Fig.~\ref{BP1BP2DM}~(c) and (d) show $\tilde{\sigma}_{\mathrm{SI}}$ as a function of $m_\chi$ at BP1 and BP2, respectively. The dotted line in each figure represents the bound from the LZ experiment. 
The model prediction in both figures is allowed around at $m_\chi$= 62.5 GeV, which is affected by the dips in $\Omega_{\chi}h^2$. In each figure, although the cross section at $m_\chi \simeq 10~\mathrm{TeV}$ satisfies the limit from the experiment, it is excluded by the constraint on $\Omega_{\mathrm{DM}}h^2$. Thus, in the following study, we fix the DM mass at $m_\chi= 62.5~\mathrm{GeV}$ in both BP1 and BP2. We should mention why even BP1 with the degenerate scalar scenario, where the DM-quark scattering is suppressed, can only satisfy the DM direct detection results in some regions. The mass difference between two Higgs bosons is represented by $\delta_2$~(\ref{del2}). In the degenerate scalar scenario, $\delta_2$ becomes necessarily small because the mass difference is small. However, as mentioned previously, large $\delta_2$ is necessary to realize the tree-level MPP. This conflicting requirement for $\delta_2$ has led to the situation~(for a detailed study, see Ref.~\cite{Cho:2021itv}). However, it is also true that there is an allowed region around $m_\chi$= 62.5 GeV.

Next, we discuss constraints on benchmark points from the Higgs search experiments at the LHC. 
The signal strength $\mu$ is used to compare the branching ratio of the $125~\mathrm{GeV}$ scalar at the LHC and the SM prediction. 
The LHC Run-2 experiment gives constraints on $\mu$ as 
$0.92<\mu<1.20 $ at ATLAS~\cite{ATLAS:2019nkf} and  $0.90<\mu<1.16$ at CMS~\cite{CMS:2020gsy}. 
The total decay width of the Higgs boson is constrained as $\Gamma_{h}^{\mathrm{exp}}<$  14.4 MeV at ATLAS~\cite{ATLAS:2018jym} and $\Gamma_h^{\exp }=3.2_{-1.7}^{+2.4}$ MeV at CMS~\cite{CMS:2022ley}. 
As mentioned in Sec.\ref{sec:model}, in the degenerate scalar scenario, the sum of decay rates of $h_1$ and $h_2$ is not different from the SM prediction, so both $\mu$ and $\Gamma_h$ in BP1 are not altered from those of the SM. 
In BP2, since the second scalar mass $m_{h_2}$ is lighter than half of $m_{h_1}$, a decay channel $h_1 \to h_2 h_2$ opens, and it may affect $\Gamma_h$ (and the branching ratio). 
However, since the mixing angle $\alpha$ is chosen very small $(0.001)$ in BP2, 
the total decay width of $h_1$ is given by $\Gamma_{h_1}$=4.29 MeV, which does not much differ from the SM prediction, $\Gamma_{h}^{\mathrm{SM}}$=4.1~\cite{LHCHiggsCrossSectionWorkingGroup:2013rie}. We also find that the signal strength is $\mu=0.956$ in BP2, which is consistent with the results at ATLAS and CMS.

\begin{table}[t]
\center
\begin{tabular}{|c|c|c|c|c|c|c|}
\hline
 & $v_C/T_C$ & $v_{SC}$ [GeV] & $v_{SC}^{\prime}$ [GeV]\\ \hline
~BP1~ & $\frac{244.0}{48.3}=5.1$  & 0.62 & 214.6\\ 
~BP2~ & $\frac{244.4}{49.7}=4.9$  & 10.3  & 226.2\\  
\hline
\end{tabular}
\caption{VEVs at critical temperature $T_C$ in BP1 and BP2. The calculations are performed using \texttt{cosmoTransitions}~\cite{Wainwright:2011kj}. Thermal resummation is needed to improve perturbative expansion at a finite temperature. The Parwani resummation method~\cite{Parwani:1991gq} is used in this analysis.}
\label{tab:TCvC}
\end{table}
We have so far confirmed that two benchmark points BP1 and BP2 satisfy constraints from DM experiments (relic density and direct detection) and the Higgs search experiments at the LHC. 
Now, we discuss the numerical evaluation of EWPT. The calculations were performed using \texttt{CosmoTransitions}~\cite{Wainwright:2011kj}, and results are obtained using the 1-loop effective potential~(\ref{eff}). However, perturbation theory is broken down due to boson multi-loop contributions, which need to be addressed in thermal resummation methods. In this study, we use the Parwani resummation method~\cite{Parwani:1991gq} in which all the Matsubara frequency modes are resummed. Specifically, field-dependent masses $\bar{m}_i^2$ that appears in $I_{B,F}$ (\ref{finite}) are replaced by thermally corrected field-dependent masses.

Table~\ref{tab:TCvC} shows $T_C$ and the corresponding VEVs in two benchmark points. Note that since the parameters for which the tree-level MPP is valid are chosen, no first-order EWPT derived from the structure of the tree-level potential occurs. However, in the CxSM, thanks to additional bosons, strong first-order EWPT could be achieved at the 1-loop level with a finite temperature. In fact, $v_C/T_C=5.1$ for BP1 and $v_C/T_C=4.9$ for BP2, satisfying the conditions for strong first-order EWPT~(\ref{decoupling})\footnote{A distinctive feature of this EWPT is sizable change in $v_{S}$, i.e., $v_{SC}^{\prime}\gg v_{SC}$. For the devoted study, see Ref.~\cite{Cho:2021itv}.}.

\section{Summary}\label{sec:Summary}

In this paper, we have applied the tree-level MPP to the CxSM, which includes a linear term of singlet $S$ to avoid the domain-wall problem. 
In the CxSM, it is known that when a mass of an additional scalar is approximately degenerate with the SM Higgs mass, the DM-quark scattering amplitudes mediated by two scalars are canceled,   and constraints on the model from the DM direct detection experiments are significantly weakened~\cite{Abe:2021nih}. 
On the other hand, the tree-level MPP chooses parameters so that multiple vacua in the models with the extended scalar sector have degenerate energy density. 
We discussed the possibility that the model parameter space to realize such a degenerate scalar scenario is favored from the tree-level MPP by requiring the electroweak vacuum ($v,v_S$) and the singlet vacuum ($0,v_S^{\prime}$) are degenerate. 

The possible existence of parameters that eliminate the potential difference in two vacua (\ref{potentialdifference}) has been investigated in the degenerate scalar region and the non-degenerate scalar region for comparison. A degeneracy between two vacua requires a difference between $v_S$ and $v_S^{\prime}$, and $v_S$ must be small in both regions. We found the parameter space where two vacua are degenerate with small $v_S$ (Fig.\ref{degenerateMPP},\ref{nondegenerateMPP}). 

We have considered two benchmark points that satisfy the tree-level MPP requirement, in which the second scalar mass is fixed at $m_{h_2}=124$ GeV for BP1 and $m_{h_2}=10$ GeV for BP2. Our numerical analysis showed only the DM mass $m_\chi \simeq 62.5~\mathrm{GeV}$ is consistent with the DM relic density observation and the DM direct detection experiment for both benchmark points. We also have discussed the feasibility of first-order EWPT. The critical temperature $T_C$ for EWPT is the temperature at which the effective potential has two degenerate minima. Thus the tree-level MPP conditions are very similar to the tree-level driven EWPT as in the CxSM, where the tree-level potential provides the leading contribution~[see Eqs.~(\ref{TCdefinition}),(\ref{MPPdefinition})]. We pointed out that the tree-level contribution to the EWPT is incompatible with the tree-level MPP. However,  it is numerically shown that the subleading contribution from additional bosons at the 1-loop level causes strong first-order EWPT~[see Table.\ref{tab:TCvC}].


\begin{acknowledgments}
We are grateful to Eibun Senaha for his valuable discussions. The work of G.C.C. is supported in part by JSPS KAKENHI Grant No. 22K03616. 

\end{acknowledgments}


\bibliography{refs}
\end{document}